\documentclass[aps,prd,a4paper,eqsecnum,groupedaddress,showpacs,
twocolumn,
showkeys,amsfonts,amssymb,amsmath,epsfig]{revtex4-1}

\usepackage{graphicx}  
\usepackage{psfrag}
\usepackage{pstricks}
\usepackage{epsf}
\usepackage{amssymb,latexsym}
\usepackage[all]{xypic}

\usepackage[mathscr]{eucal}
\usepackage{citesort}
\usepackage[active]{srcltx}
\usepackage{url}
\usepackage{epstopdf}

 \newcommand{\tg}{{\tilde{g}}}
\newcommand{\ztheta}{\mathring \theta}

\newcommand{\mytsigmasquare}{\tilde \sigma^2}
\newcommand{\tbeta}{\tilde\beta}
\newcommand{\nthree}{\nu}
\newcommand{\psiHDS}{\psi}
\newtheorem{theorem}{Theorem}[section]
 \newcommand{\bean}{\begin{eqnarray}\nonumber}
\newcommand{\beal}[1]{\begin{eqnarray}\label{#1}}
\DeclareFontFamily{OT1}{rsfs}{}
\DeclareFontShape{OT1}{rsfs}{m}{n}{ <-7> rsfs5 <7-10> rsfs7 <10-> rsfs10}{}
\DeclareMathAlphabet{\mycal}{OT1}{rsfs}{m}{n}
\newcommand{\R}{{\mathbb R}}

\newcommand{\rd}{\,{ d}} 
\newcommand{\tL}{{\widetilde L}}
\newcommand{\tR}{{\widetilde R}}
\newcommand{\Eq}[1]{Equation~\eq{#1}}
\newcommand{\be}{\begin{equation}}
\newcommand{\eeq}{\end{equation}}
\newcommand{\ee}{\end{equation}}
\newcommand{\beqa}{\begin{eqnarray}}
\newcommand{\eeqa}{\end{eqnarray}}
\newcommand{\beqan}{\begin{eqnarray*}}
\newcommand{\eeqan}{\end{eqnarray*}}
\newcommand{\ba}{\begin{array}}
\newcommand{\ea}{\end{array}}
\newcommand{\const}{\mbox{\rm const}} 
\newcommand{\hyp}{\mycal S}
\newcommand{\mcM}{{\mycal M}}
\newcommand{\mnote}[1]
{\protect{\stepcounter{mnotecount}}$^{\mbox{\footnotesize
$
\bullet$\themnotecount}}$ \marginpar{
\raggedright\tiny\em
$\!\!\!\!\!\!\,\bullet$\themnotecount: #1} }

\newcommand{\eq}[1]{(\ref{#1})}
\newcommand{\mcL}{{\mycal L}}
\newcommand{\beaa}{\begin{eqnarray*}}
\newcommand{\eeaa}{\end{eqnarray*}}

\newcommand{\zR}{\mathring{R}}

\begin{document}

\preprint{Preprint UWThPh-2013-2}
\title{Hamiltonian mass of asymptotically Schwarzschild-de Sitter space-times}

\author{Piotr T. Chru\'{s}ciel}
\affiliation{Universit\"at Wien, Gravitational Physics, Boltzmanngasse 5, A1090 Wien, Austria}
    \author{Jacek Jezierski}
\affiliation{Uniwersytet Warszawski, Wydzia{\l} Fizyki, KMMF, ul.\ Ho\.za 74, PL 00-681 Warszawa, Poland}
\author{Jerzy Kijowski}
\affiliation{Center for Theoretical Physics, Polish Academy of Sciences, al.\ Lotnik\'ow 32/46, PL 02-668 Warszawa, Poland}

\date{\today}

\begin{abstract}
We derive the Hamiltonian mass for general relativistic initial data sets with asymptotically Schwarzschild-de Sitter ends.
\end{abstract}


\maketitle

\section{Introduction}

There is  growing astrophysical evidence that space-times with positive cosmological constant should be given serious consideration.
Large families of such non-compact, vacuum, general-relativistic models can be constructed using singular solutions of the Yamabe problem (see~\cite{ChPollack,CPP} and references therein). In particular one thus obtains initial data sets with one or more ends of cylindrical type, in which the metric becomes periodic  when one recedes to infinity along  half-cylinders,
approaching the Schwarzschild-de Sitter metric in the limit, with the extrinsic curvature tensor approaching zero.
This construction can be carried out in any number of space dimensions $n\ge 3$. (See~\cite{CortierKdS,CMP,CM,BMW,GabachClement} for further families of vacuum initial data sets with various ends of cylindrical type.) This raises the question of existence of a natural notion of mass in this context. The object of this work is to show that the numerical value of a natural Hamiltonian $ {\mathcal H} $ for a class of such metrics is proportional to the parameter $m$ appearing in the asymptotic metric. We further prove that the contribution to the Hamiltonian
from each asymptotically Schwarzschild-de Sitter end  can be calculated as
%
\begin{equation}\label{16XII12.8xa}
 {\mathcal H} = \lim_{x_0 \to
 \infty} \frac 1{{ 2 \gamma}} \int_{x=x_0}  (k \nu - k_0\nu_0)\lambda \,d^{n-1}x 
 \;.
\end{equation}
Here we assume that the space-metric is asymptotic to the space-part of a Birmingham metric on $[0,\infty)\times \mathring M$  as in  \eq{6XI12.4}-\eq{6XI12.5}, for a compact Riemannian $(n-1)$-dimensional Einstein manifold $(\mathring M,\mathring h)$; $\nu$ is the  lapse function as in \eq{n,nA}; $k$ is the mean curvature of $\{x=x_0\}$ as defined in \eq{1III13.1}; and $\lambda$ is the $(n-1)$-volume element on $\{x=x_0\}$.
The
fields $\nu_0$ and $k_0$ are the corresponding quantities for the Birmingham metric with vanishing mass (the ``de Sitter solution"), see \eq{16XII12.7a}-\eq{11IV13.1}.
Finally $\gamma$ is a dimension-dependent coupling constant, see \eq{3I13.7} in Appendix~\ref{s3I13.1} below, related to the ``$(n+1)$-dimensional Newton constant" as in \eq{3I13.11}.

We note that a Hamiltonian is always defined up to a constant. Our choice in \eq{16XII12.8xa} is precisely what is needed for positivity of $\mathcal H$,  compare Theorem~\ref{t2I13.1} below.

See~\cite{AbbottDeser,AnninosMusings,BalaBD} and references therein for alternative approaches to a definition of mass in the presence of a positive cosmological constant.

\section{The basic variational formula}
 \label{s16IV13.1}

In order to present our results some notation is needed. Let $\hyp$ be a smooth spacelike hypersurface in an
 $(n+1)$-dimensional space-time $(\mcM,g)$, $n\ge 2$.
Consider a space-time domain $\Omega$ with smooth timelike boundary such that  $V:=\Omega\cap \hyp$ is compact.
Let $x^n$ be a coordinate such that $x^n$ is constant on $\partial V$, and let $(x^a)=(x^0, x^A)$ be local coordinates on $\partial \Omega$ such that $x^0$ is constant on $\hyp$. Let $L_{ab}$ denote the extrinsic curvature tensor of $\partial \Omega$,
\begin{eqnarray}
L_{ab} & = & - \frac 1{\sqrt{g^{nn}}} {\Gamma}^n_{ab}\; ,
\end{eqnarray}
and let $Q^{ab}$ be its ``ADM counterpart",
\begin{eqnarray}
Q^{ab} & := & \sqrt{|\det g_{cd}|} \ (L {\hat g{}}^{ab} - L^{ab} ) \; ,
\label{4I13.6}
\end{eqnarray}
where ${\hat g{}}^{ab}$ is the $n$-dimensional inverse with respect to the
induced metric $g_{ab}$ on the world-tube $\partial \Omega$.
Let $\nthree$
and $\nthree^A$ denote the ``lapse'' and the ``shift'' in the
$n$-dimensional geometry $g_{ab}$ of the boundary of the world-tube $\partial \Omega$,
\begin{equation}
\nthree := \frac 1{\sqrt{|{\hat g{}}^{00}|}}\; ,  \ \ \ \ \ \ \ \
\nthree^A := {\tilde{\tilde g}}^{AB} g_{0B} \; , \label{n,nA}
\end{equation}
where ${\tilde{\tilde g}}^{AB}$ is the $(n-1)$-dimensional metric on $\partial
V$, inverse with respect to the induced metric $g_{AB}$. We have the identity
\begin{equation}
g_{00} = - \nthree^2 + \nthree_A \nthree^A \;. \label{g00}
\end{equation}
One can define
the following $(n-1)$-dimensional objects on $\partial V$:
a scalar density
\begin{equation}
 \label{15XII12.1}
{\bf Q}  :=  \nthree Q^{00}
\; ,
\end{equation}
and a covector density
\begin{equation}
{\bf Q}_A  :=  Q^0_{\ A} \;.
\end{equation}
It is further useful to introduce  the  field
\begin{equation}
\stackrel{\perp}{Q}\!{^{AB}}  :=  Q_{CD} \, {\tilde{\tilde g}}^{CA}
\, {\tilde{\tilde g}}^{DB} \;.
\end{equation}
The $n$-dimensional Lorentzian metric $g_{ab}$ on $\partial \Omega$ can be parameterized as
\begin{equation}\label{gd}
{g}_{ab} = \left[ \begin{array}{ccc}
-\nu^2+ \nu^A \nu_A & \vline & \nu_A
\\[0.7ex]
\hline
 \nu_A & \vline & g_{AB}  \\
 \end{array} \right] .
\end{equation}
The corresponding inverse metric reads
\begin{equation}\label{gg}
{\hat g{}}^{ab} = \left[ \begin{array}{ccc}
-\frac1{\nu^2} & \vline & \frac{\nu^A}{\nu^2} \\[1ex]
\hline
{\phantom{\displaystyle\int}}\frac{\nu^A}{\nu^2} & \vline & {\tilde{\tilde g}}^{AB} - \frac{{\nu^A}{\nu^B}}{\nu^2}
 \end{array} \right] .
\end{equation}
We also have
\begin{equation}
  L^{00} = L_{ab}{\hat g{}}^{0a} {\hat g{}}^{0b} =
  \frac{1}{\nu^4} \left(L_{00}  -2 L_{0A} \nu^A + L_{AB} \nu^A \nu^B
  \right)
   \; ,
\end{equation}
with the trace $L$ of $L_{ab}$ being equal to
\begin{align*}
  L = &\,  L_{ab}{\hat g{}}^{ab} = L_{00}{\hat g{}}^{00}
  +2 L_{0A} {\hat g{}}^{0A} + L_{AB} {\hat g{}}^{AB}\\
   = & - \frac 1{\nu^2} L_{00} + 2 L_{0A} \frac{\nu^A}{\nu^2} +
   L_{AB}  \left(
   {\tilde{\tilde g}}^{AB} - \frac{{\nu^A}{\nu^B}}{\nu^2}
   \right)
    \\
  = & - \nu^2 L^{00} + L_{AB} {\tilde{\tilde g}}^{AB}
  =   a- k
      \; ,
\end{align*}
where
\[
    k := - L_{AB} {\tilde{\tilde g}}^{AB}
\]
(for the Birmingham metrics of Appendix~\ref{ss7I13.11}, $k$ is the signed length of the extrinsic curvature vector),
and where we use the symbol
\[
    a :=  - \nu^2 L^{00}
\]
to denote the curvature (``acceleration")
of the world-lines which are geodesic within
$\partial \Omega$ and orthogonal to $\partial \Omega \cap \hyp$.
It holds that
\begin{eqnarray}
 \nonumber
    {\bf Q}
     &= &
       \nu Q^{00} = \nu^2 \lambda \left(
    L {\hat g{}}^{00} - L^{00}
    \right)
\\
 \label{1III13.1}
     & =  &  \lambda \left(
    -L - \nu^2 L^{00}
    \right)=
     \lambda k \; ,
\\
 \nonumber
 \stackrel{\perp}{{Q}}\!{^{AB}} g_{AB} &=&  Q_{CD} \, {\tilde{\tilde g}}{^{CA}}
 \, {\tilde{\tilde g}}{^{DB}} g_{AB} = Q_{CD} {\tilde{\tilde g}}{^{CD}}
\\
\nonumber
 & = &
 \nu \lambda \left( L g_{AB} - L_{AB}
 \right) {\tilde{\tilde g}}{^{AB}}
  \\ &=&
    \nu \lambda \left( (n-1) a  - (n-2) k
\right)
\;.\label{1III13.2}
\end{eqnarray}

Let  $P^{ij}$ be the usual ADM momentum on $V$. Denote by
$$
 \lambda=\sqrt{\det g_{AB}}
$$
the $(n-1)$-volume element on $\partial V$. Let $\alpha$ be the hyperbolic angle between $\partial \Omega$ and
$V$: in the adapted coordinates above,
$$
 \alpha := \sinh ^{-1}\left( \frac {g^{0n}}{\sqrt{|g^{00} g^{nn}|}}\right)
 \;.
$$
In~\cite{KijowskiGRG} the following variational formula has been proved for Ricci-flat Lorentzian metrics in dimension $3+1$,
\begin{eqnarray}
0 & = & \frac 1{{ 2 \gamma}} \int_V  \left( {\dot P}^{kl}  \delta g_{kl} -
{\dot g}_{kl} \delta P^{kl} \right) +
\frac 1{{  \gamma}} \int_{\partial V} ( {\dot \lambda} \delta \alpha  -
{\dot \alpha} \delta \lambda )
\nonumber \\
 &   & +\frac 1{{ 2 \gamma}} \int_{\partial V} (  2 \nthree\delta {\bf Q}
- 2\nthree^A \delta {\bf Q}_A
+ \stackrel{\perp}{Q}\!{^{AB}}
 \delta g_{AB} ) \; ,  \label{homogeneous}
\end{eqnarray}
with $\gamma=8\pi$. It can be checked that the formula remains true for vacuum metrics, possibly with a cosmological constant, in any space-dimension $n\ge 2$,
 with a constant $\gamma $
which depends upon dimension; see Appendix~\ref{s3I13.1} for a discussion.
In fact, several terms proportional to $(n-3)$ appear when generalizing the calculations in~\cite{KijowskiGRG}, but they end-up giving no contribution to  \eq{homogeneous}.

We will not dwell upon the Hamiltonian interpretation of this identity, the reader is referred to~\cite{KijowskiGRG,KijowskiTulczyjew,CJK} for details.

In the non-vacuum case  \eq{homogeneous} has to be supplemented by terms involving variations of the matter fields and their  momenta.
Nevertheless, the formula \eqref{16XII12.8xa} for the Hamiltonian remains valid for a large class of matter models~\cite{KijowskiGRG} without any further explicit contributions from the matter sources. (Obviously, there is an implicit contribution of the sources via the constraint equations.)

\section{The mass of asymptotically-Birmingham metrics}
 \label{s3III13.1}

We consider \eq{homogeneous} for metrics which, as $x$ tends to infinity, asymptote to
\begin{equation}\label{10XII12.1}
 \mathring g= - f(x) dt^2 + \phi^2(x) (dx^2 + \mathring h_{AB}dx^ A dx^B )
 \;.
\end{equation}
Similarly we will assume that the derivatives of the metric $g$ asymptote to those of the metric $\mathring g$. The coordinate $x^n$ of the calculations above will be taken to be equal to $x$, and the boundary $\partial V\approx \mathring M $ in \eq{homogeneous} will be assumed to be
given by the equation $x=x_0$ for a constant $x_0$. We will let $x_0$ tend to infinity; this implies
\bean
   L_{ab}dx^ a dx^ b & = &   - \frac 1{\sqrt{g^{xx}}} {\Gamma}^x_{ab}dx^ a dx^ b
\\ \label{12XII12.1}
 &  \displaystyle \rightarrow  & - \frac 12 \phi^{-1}\partial_x f dt^2 + \partial_x \phi \, \mathring h_{AB}dx^ A dx^ B
  \; ,
  \phantom{xx}
\end{eqnarray}
\begin{eqnarray}
   L &=& g^{ab}L_{ab} \displaystyle \rightarrow    \frac{ \partial_x f}{2\phi f}   + (n-1) \frac{\partial_x\phi}{\phi^2}
   \; ,
\\
 a & = &    - \nu^2 L^{00} \displaystyle \rightarrow  \frac{ \partial_x f}{2\phi f}
  \; ,
\\
 \nthree & \to &   \sqrt f
 \; ,
 \quad
  \lambda  \to\phi^{n-1} {\sqrt{\det {\mathring h_{AB}}}}
 \; ,
 \quad
  \nthree_A \to 0
 \; ,
\end{eqnarray}
\begin{eqnarray}
{\bf Q} & = &  \nu Q^{00} \to - (n-1) \phi^{n-3} {\sqrt{\det {\mathring h_{AB}}}}\partial_x \phi
 \; ,
\\
 k  & = & \lambda^{-1}{\bf Q} \to - (n-1) \phi^{-2}  \partial_x \phi
 \; ,
\\
 {\bf Q}_A  &  \to & 0
  \; ,
\\ 
\stackrel{\perp}{{Q}}\!{^{AB}}   & \to & \sqrt{f} \phi^{n-3}{\sqrt{\det {\mathring h_{AB}}}}
  \left( \frac{ \partial_x f}{2\phi f}   + (n-2) \frac{\partial_x\phi}{\phi^2}\right)\mathring h^{AB}
   \nonumber
\\
 & = &   \sqrt{f} \phi^{n-4}{\sqrt{\det {\mathring h_{AB}}}}
   \frac{ \partial_x (\sqrt f \phi ^{n-2})}{\sqrt f \phi ^{n-2}}  \mathring h^{AB}
 \; .\label{12XII12.2}
\end{eqnarray}
Above, and in what follows, we assume that $\partial_x$ is pointing outwards from the region $V$ of the previous section; some signs adjustments are needed otherwise.
Using these formulae, the last line in \eq{homogeneous} approaches
\begin{equation}\label{15XII12.11}
  \frac {(n-1)}{{ 2 \gamma}} \int_{\mathring M}  \sqrt{f} \phi^{n-3}{\sqrt{\det {\mathring h_{AB}}}}
  \left( \frac{ \partial_x (f \phi^2)}{\phi^2 f}   \delta\phi - 2 \partial_x \delta \phi\right)
   \;.
\end{equation}

Let us assume that $f$ and $\partial_x \phi$ take the Birmingham form \eq{6XI12.4}%
\footnote{In this work we consider only asymptotically time-symmetric models, and therefore only
the solutions discussed in Section~\ref{ss7I13.11}: The exactly cylindrical solutions of
Appendix~\ref{ss6I13.1}
are not time-symmetric, while the solutions of Appendix~\ref{ss6I13.2} have no asymptotically cylindrical regions.}%
,
\begin{equation}\label{15XII12.13}
 f= \beta -
 \frac {2m} {\phi^{n-2}} - \frac{\phi^2}{\ell^2}
 \; ,
  \quad \partial_x \phi = \phi \sqrt{f}
  \; ,
\end{equation}
where $\beta\in\{0,\pm 1\}$ is related to the scalar curvature,  assumed to be constant, of the metric $
\mathring h$ (see \eq{4I13.4}). Finally, $\ell^{-2}$ is related to the cosmological constant as in \eq{17XII12.21}.
When $\mathring h$ is the unit round metric on the sphere, then $\beta=1$ and one recovers the familiar Schwarzschild-de Sitter metrics.
 \Eq{15XII12.13} allows us to express $\partial_x \delta \phi = \delta \partial_x \phi$   in terms of $\delta \phi$ and $\delta m$.
Perhaps surprisingly, all the $\delta \phi$ terms cancel out and \eq{15XII12.11} becomes
\begin{equation}\label{15XII12.12}
  \frac {(n-1)}{{  \gamma}} \int_{\mathring M}  {\sqrt{\det {\mathring h_{AB}}}}
  \times
  \delta m
   \;.
\end{equation}
Setting
\begin{equation}\label{15XII12.4}
 {\mathcal H} =
  \frac {(n-1)}{{  \gamma}} \int_{\mathring M}  {\sqrt{\det {\mathring h_{AB}}}}
  \times
   m
   \; ,
\end{equation}
we conclude that for any family of metrics which asymptote to Birmingham metrics as the variable $x$ recedes to infinity
it holds that
\begin{eqnarray}
 - \delta {\mathcal H} & = & \frac 1{{ 2 \gamma}} \int_\hyp  \left( {\dot P}^{kl}  \delta g_{kl} -
 {\dot g}_{kl} \delta P^{kl} \right)
 \;.
\label{15XII12.21}
\end{eqnarray}
This is the first main result of this work.

We wish, next, to provide a geometric formula for the Hamiltonian $\mathcal H$.
The integrand of the boundary term in \eq{homogeneous},
\begin{equation}\label{1III13.4}
     2 \nthree \delta {\bf Q}
- 2 \nthree^A \delta {\bf Q}_A
+ \stackrel{\perp}{{Q}}\!{^{AB}}
 \delta g_{AB}
  \; ,
\end{equation}
can be rearranged using the identity
\begin{equation}\label{1III13.3}
    2\nthree\delta {\mathbf Q} = 2\nthree\delta (\lambda k) =
    \delta(\lambda \nthree k) + \lambda \nu^2 \delta (\frac k\nu ) + \nu k \delta \lambda
     \;.
\end{equation}
Using
$$
 \nu k \delta \lambda = \frac 12 \lambda \nu k \tilde{\tilde g}^{AB} \delta g_{AB}
$$
we can write
\begin{equation}\label{16XII12.2}
  2 \nu \delta {\bf Q}
+ \stackrel{\perp}{{Q}}\!{^{AB}}
 \delta g_{AB}
   = \delta(\lambda \nu k) +
    \underbrace{\lambda \nu^2 \delta (\frac  k\nu ) +
  {\bf Q}^{AB}
 \delta g_{AB}}_{(*)}
  \; ,
\end{equation}
where
\begin{equation}
\label{Q-bold}
    {\bf Q}^{AB} :=
  \; \stackrel{\perp}{Q}\!{^{AB}}
 + \frac 12 \lambda \nu   k {\tilde{\tilde g}}^{AB} \;.
\end{equation}

As before,  we assume that the metric asymptotes to a Birmingham metric as $x$ tends to infinity, similarly for first derivatives.
We then have
\beal{16XII12.3}
 & \displaystyle
 \frac k \nu + \frac{(n-1)}{\phi} \; \raisebox{-1ex}{$\overrightarrow{\vphantom{x}\scriptstyle \,x\to\infty\,}$} \; 
 0
 \; ,
 &
\\& \nonumber
  \displaystyle
  \stackrel{\perp}{{Q}}\!{^{AB}} - \nu \phi^{n-3} \bigg( a - \frac {n-2}{n-1}k\bigg) \sqrt{\det \mathring h_{CD}} \, \mathring h^{AB} \; \raisebox{-1ex}{$\overrightarrow{\vphantom{x}\scriptstyle \,x\to\infty\,}$} \; 0
 \; ,
 &
 \\& \nonumber
  \displaystyle
  {\bf{Q}}^{AB} - \nu \phi^{n-3}  \bigg( a - \frac {n-3}{2(n-1)}k\bigg)   \sqrt{\det \mathring h_{CD}} \, \mathring h^{AB}
  \; \raisebox{-1ex}{$\overrightarrow{\vphantom{x}\scriptstyle \,x\to\infty\,}$} \; 0
 \; ,
 &
\\&
  \displaystyle
 a -  \frac {\partial_\phi f}{2\nu}
 \; \raisebox{-1ex}{$\overrightarrow{\vphantom{x}\scriptstyle \,x\to\infty\,}$} \; 0
  \;.
 &\label{16XII12.5}
\end{eqnarray}
Inserting those relations into the underbraced terms in \eq{16XII12.2} one finds
\bean
 \lefteqn{
 (*)
 \; \raisebox{-1ex}{$\overrightarrow{\vphantom{x}\scriptstyle \,x\to\infty\,}$} \;
 (n-1)\phi^{n-3} \sqrt{\det \mathring h_{CD}}  \left[ \phi \partial_\phi f +  (n-2) f \right] \delta \phi
 }
 &&
\\
 &  &\phantom{xxx} = (n-1)\delta \left[\sqrt{\det \mathring h_{CD}}
  \left( \beta \phi^{n-2} - \frac{1}{\ell^2}\phi^{n}\right) \right]
 \;.
  \phantom{xxxx}\label{11IV13.2}
\end{eqnarray}
We thus obtain the following formula for the Hamiltonian:
\begin{equation}\label{16XII12.7}
 {\mathcal H} = \lim_{x_0 \to
 \infty} \frac 1{{ 2 \gamma}}  \int_{x=x_0} \left(\nu k + (n-1) \left( \frac{\beta}{\phi } - \frac{\phi}{\ell^2} \right)
 \right)\lambda \, d^{n-1}x
 \;.
\end{equation}
Here $\phi$ should be viewed as a function of $\lambda$, hence of the metric:
\begin{equation}\label{16XII12.7a}
 \phi = \bigg(\frac{\lambda}{\sqrt{\det \mathring h_{AB}}}\bigg)^{\frac 1 {n-1}} = \bigg(\frac{\sqrt{\det g_{AB}}}{\sqrt{\det \mathring h_{AB}}}\bigg)^{\frac 1 {n-1}}
 \;.
\end{equation}

Choose a constant $m_0\in \R$ and set
\begin{equation}\label{11IV13.1}
     f_0 = \beta - \frac{2m_0}{\phi^{n-2}}-  \frac{\phi^2}{\ell^2} \; ,
     \   \nu_0 = \sqrt{f_0} \; ,
     \
    k_0 = -\frac {(n-1)}\phi \sqrt{f_0} \; .
\end{equation}
This leads to the following rewriting of \eq{16XII12.7}:
\begin{eqnarray}
 {\mathcal H} &= &\lim_{x_0 \to
 \infty} \frac 1{{ 2 \gamma}} \int_{x=x_0}  (k \nu - k_0\nu_0)\lambda \,d^{n-1}x
 \nonumber
\\
 &&  +
 \frac {(n-1)|\mathring M|_{\mathring h}}{{  \gamma}}m_0
 \; ,
 \label{16XII12.8}
\end{eqnarray}
where $|\mathring M|_{\mathring h}$ is the volume of the set $\{x=x_0\}$
 in the metric $\mathring h$. This is the second main result of this work.

Note that the parameter $m_0$ has been introduced only to define the reference fields $k_0$ and $\nu_0$, and that the left-hand side is independent of $m_0$.

See Appendix~\ref{s17XII12.11} for an alternative derivation of \eq{16XII12.8}.

One can simply disregard the last term in \eq{16XII12.8}, or use reference fields associated with the solution equal to $m_0=0$ there. Here one should keep in mind that a Hamiltonian analysis always defines a Hamiltonian up to a constant, and the choice of this constant is equivalent to the decision, which field configuration (if any) has zero energy. As such, the subtraction of the term $k_0\nu_0$ can be viewed as a comparison term, where one compares the given field configuration with that time-independent solution which is determined by the $m_0$-parameter.

One could argue that reference fields corresponding to the solution with $m_0=0$ make no sense because, in the $\mathring M=S^{n-1}$ case, the initial data surface is \emph{compact}, so comparing with a solution with asymptotically periodic ends is unnatural from a Hamiltonian perspective. However, one can adopt the point of view that energy in general relativity is not assigned to a \emph{volume} $V$ but rather to a \emph{surface} $\partial V$. Given   a level set of $r$ in a Schwarzschild-de Sitter solution, we can find a surface with identical induced metric in the de Sitter solution%
\footnote{Indeed, both  extremal values $r_0^{(1)}$ and $r_0^{(2)}$ of Figure \ref{F6XI12.1} belong to the interval $[0,\ell]$, see \eq{13IV13.1}. }%
,
$m_0=0$, and use the corresponding values $\nu_0$ and $k_0$  in  \eq{16XII12.8}.

In any case, somewhat surprisingly, the choice of the  value of $m_0$ is irrelevant, in that  the numerical value of $\mathcal H$ as given by \eq{16XII12.8} does not depend upon that choice. This is related to the fact that the mass parameter $m$ is the (unique) ``constant of motion'' for the spherically-symmetric Yamabe equation, cf.~\eq{6XI12.13}.

We note that the time-symmetric Birmingham metrics lead to  the periodic metrics \eq{10XII12.1} with a strictly positive parameter $m$, see the discussion in Appendix~\ref{s4I13.1}. This leads to the following trivial observation:

\begin{theorem}[``Positive energy theorem"]
 \label{t2I13.1}
For all asymptotically periodic metrics as above, the numerical value of the Hamiltonian $\mathcal H$
  given by \eq{16XII12.8} is positive.
\end{theorem}

Now, Theorem~\ref{t2I13.1} does neither require positivity of matter-energy  nor regularity of initial data  (in particular interior boundaries are allowed without any geometric restrictions), and is based purely on asymptotic properties of the solutions. As such it does not carry much non-trivial information: the positivity of the mass has been built-in into the hypotheses on the asymptotic behavior of the metric.

\subsection{Several ends, black hole boundaries}

So far we have assumed that the initial-data manifold is the union of a compact manifold without boundary and an asymptotically cylindrical end. The generalization of our analysis to  a finite number of asymptotically flat, asymptotically cylindrical and asymptotically hyperboloidal ends is straightforward: In such a case each end contributes
its respective Hamiltonian mass (as defined here for asymptotically Birmingham ends, and as defined in e.g.~\cite{CJL,ChAIHP,CJK,BOM:poincare} and references therein for the remaining ones) to the total Hamiltonian of the system.

Yet another generalization is of interest, that to manifolds with \emph{horizon boundaries}. For this purpose, suppose that the boundary of the domain $\Omega$ of Section~\ref{s16IV13.1} consists of a timelike ``world tube" $S^+ $ and of a null hypersurface $S^-$.
Accordingly, the boundary $\partial V$ of the  intersection $V$ of the Cauchy surface $\hyp$ with $\Omega$ is composed of two disjoint manifolds, $\partial V^+=
V\cap S^+$, and $\partial V^-=V\cap S^-$, assumed to be compact, each of them contributing to the boundary terms in variational formula \eq{homogeneous}. Assume that the space-time metric  asymptotes to a Birmingham metric as the ``external'' boundary $\partial V^+$ recedes to infinity. The corresponding contribution to \eq{homogeneous} is handled as in the previous section.
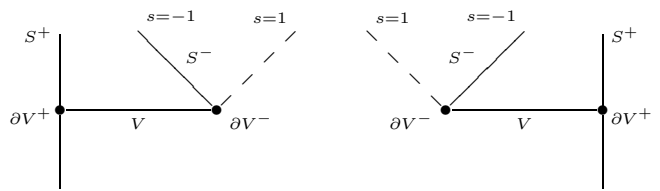
\begin{figure}[h]
\[%
\xymatrix@M=0pt@R=0.95cm@C=0.95cm{ \ar@{-}[d]_<{S^+} &
\ar@{-}[dr]^<{s=-1}^{S^-}& & \ar@{--}[dl]^>{\partial V^-
}_<{s=1}&\ar@{--}[dr]_>{\partial V^- }^<{s=1}&&
\ar@{-}[dl]_<{s=-1}_{S^-} &\ar@{-}[d]^<{S^+}\\
{\bullet}\ar@{-}[r]_>V& \ar@{-}[r]&{\bullet}&&&{\bullet}\ar@{-}[r]
& \ar@{-}[r]_<V&{\bullet}\\ \ar@{-}[u]^>{\partial
V^+}&&&&&&&\ar@{-}[u]_>{\partial V^+} }\]
\caption{\label{horinfty}The orientation of $\partial V^-$.}
\end{figure}
The  contribution to \eq{homogeneous} from the null component $S^-$ was calculated in \cite{CJKbhthermo} in considerable generality. However, for the sake of simplicity,  we restrict attention to  stationary solutions with Killing horizons, as  arising in a thermodynamical  analysis
of stationary black holes. Then the volume term in \eq{homogeneous} vanishes identically  (since the time derivatives vanish) and the entire formula reduces to (see \cite[Equation~4.2]{CJKbhthermo})
\begin{equation}
 \label{form-zerowa11}
 \delta {\mathcal H}  =
    \frac s{{  \gamma}}
  \int_{\partial V^-} \left(\kappa\delta\lambda + \nu^A\delta
  {\cal W}_A \right)\; ,
\end{equation}
where the right-hand side is the (only remaining) boundary term%
\footnote{The formula has been proved in dimension three in~\cite{CJKbhthermo}, but remains true in higher dimensions.}
corresponding to the cross-section $\partial V^-$ of the horizon $S^-$.
Here $\mathcal H$ is our Hamiltonian \eq{16XII12.8}, $s=\pm 1$ is a constant which depends upon the time-orientation of the Killing vector so that   $-s \kappa$ is the surface gravity  in usual circumstances (one should also keep in mind further negative signs in \eq{form-zerowa11} which might arise from the orientation of the boundary,  see Figure \ref{horinfty}).
The field ${\cal W}_A$ is defined on the horizon by the  formula
\[ {\cal W}_A  =  -\lambda dx^0(\nabla_A K)  \; , 
  \]
  where $K$ is a Killing vector field which is null on a horizon, assuming that the horizon is located at $x^n=\const$,  and  that $x^0$ is a coordinate on the horizon satisfying
\[
\quad dx^0(K)=1
 \;.
\]

It is conceivable that the only such vacuum black-hole space-times which are asymptotic to the Birmingham metrics are the Birmingham metrics themselves, in which case the ``thermodynamical identity" \eq{form-zerowa11} can be derived by the trivial calculation  of Appendix~\ref{ss28XII12.1}. However,  this is not clear, and unlikely in higher dimensions in any case.

As already emphasized, the ``Positive Energy Theorem"~\ref{t2I13.1} remains valid in the black hole setting.

\appendix

\section{Birmingham metrics}
 \label{s4I13.1}

Consider an $(n+1)$-dimensional metric, $n\ge 3$,  of the form
\begin{equation}\label{6XI12.4}
 g = - f dt^2 + \frac {dr^2} f + r^2  \underbrace{\mathring h_{AB}(x^C) dx^A dx^B}_{=:\mathring h}
 \; ,
\end{equation}
where $\mathring h$ is a Riemannian metric on a compact manifold $\mathring M$ with constant scalar curvature $\mathring R$; we denote by $x^A$ local coordinates on $\mathring M$.  As discussed in \cite{Birmingham}, for any $m\in \R$ and $\ell\in \R^*$ the function
\begin{equation}\label{6XI12.5}
 f= \frac{\zR}{(n-1)(n-2)} - \frac {2{ m}}{r^{n-2}} - \frac{r^2}{\ell^2}
\end{equation}
leads to a vacuum metric,
\begin{equation}\label{4I13.1}
 R_{\mu\nu} =  \frac{n}{\ell^2} g_{\mu\nu}
 \; ,
\end{equation}
thus $\ell$ is a constant related to the cosmological constant as in \eq{17XII12.21} below.
(Clearly, the case $n=2$ would require separate considerations, and we will therefore ignore this dimension in our work.)
The multiplicative factor two in front of $m$ is convenient in dimension three when $\mathring h$ is a unit round metric
on $S^2$, and we will keep this factor regardless of topology and dimension of $\mathring M$.

There is a rescaling of the coordinate $r=b  \bar r$, with $b\in \R^*$, which leaves
\eq{6XI12.4}-\eq{6XI12.5} unchanged (up to ``adding bars") if moreover
\begin{equation}\label{4I13.2}
 \overline{\mathring h} = b^2\mathring h
 \; ,
 \quad
 \bar m =b^{-n} m
  \; ,
   \quad
    \bar t=  b t
    \;.
\end{equation}
We can use this to achieve
\begin{equation}\label{4I13.4}
 \beta:= \frac{\zR}{(n-1)(n-2)} \in \{0, \pm 1\}
 \; ,
\end{equation}
which will be assumed from now on. The set $\{r=0\}$ corresponds to a singularity when $m\ne 0$. Except in the case $m=0$ and $\beta = -1$,
by an appropriate choice of the sign of $b$ we can always achieve $r>0$ in the regions of interest. This will also be assumed from now on.

For reasons which should be clear from the main text, we will now be seeking functions $f$ which, after a suitable extension of the space-time manifold and metric, lead to  spatially periodic solutions.

\subsection{Cylindrical solutions}
 \label{ss6I13.1}

Consider, first, the case where $f$ has no zeros. Since $f$ is negative for large $|r|$, $f$ is negative everywhere.  It therefore makes sense to rename $r$ to $\tau>0$,  $t$ to $x$, and $-f$ to $F>0$,
leading to the metric
\begin{equation}\label{6XI12.4x}
 g =   - \frac {d\tau ^2} {F(\tau)} + F(\tau) dx^2 + \tau ^2 \mathring h
 \;.
\end{equation}
The level-sets of the time coordinate $\tau$ are infinite cylinders with topology $\R\times \mathring M$, with a product metric. Note that the extrinsic curvature of those level sets is never zero because of the $\tau^2$ term in front of $\mathring h$, except possibly for the $\{\tau=0\}$-slice in the case $\beta=-1$ and $m=0$.

\begin{figure}[t]
  \centering
    \includegraphics[scale=.85]{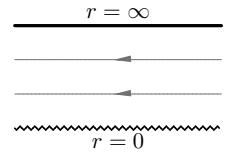}
\caption{The $(t,r)$-projection diagram when $m<0$ and $f$ has no zeros.
    \label{F28II13.1}}
\end{figure}

Assuming that $m\ne 0$, the region $r\equiv\tau\in (0,\infty)$  is  a ``big-bang -- big freeze" space-time with cylindrical spatial sections. A $(\tau,x)$-projection diagram (in the sense of~\cite{COS}) is  an infinite horizontal strip with a singular spacelike boundary at $\tau=0$, and a smooth conformal spacelike boundary at $\tau=\infty$, see Figure~\ref{F28II13.1}.

In the case $m=0$ and $\beta=0$ the spatial sections are again cylindrical, with the boundary $\{\tau=0\}$ being now at infinite temporal distance: Indeed, setting $T=\ln \tau$, when $m=0$ and $\beta=0$ we can write
\beaa
 g &=& -\ell^2 \frac {d\tau^2}{\tau^2} + \frac {\tau^2}{\ell^2} dx^2 + \tau^2 \mathring h
\\
  & = &
    -\ell^2 dT^2  + e^{2T}\left(\frac {dx^2}{\ell^2}  +   \mathring h\right)
 \;.
\end{eqnarray*}
When $\mathring h$ is a flat torus, this is one of the forms of the de Sitter metric~\cite[p.~125]{HE}.

\begin{figure}
\begin{center}
{\includegraphics[scale=.85,angle=0]{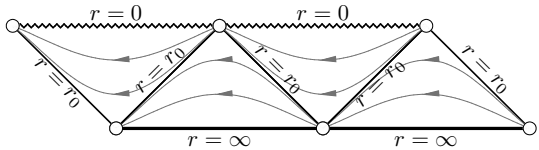}}
\caption{The $(t,r)$-projection diagram  for (suitably extended) Birmingham metrics with $f\le 0$, and $f$ vanishing precisely at $r_0$. { }}
\label{F6XI12.1x}
\end{center}
\end{figure}

The next  case  which we consider is $f\le 0$, with $f$ vanishing precisely at one positive value $r=r_0$. This occurs if and only if $\beta=1$ and
\begin{equation}\label{6I13.3}
 r_0 = \sqrt{ \frac {n-2}{n}} \ell
  \; ,
  \quad
  m= \frac{r_0^n}{(n-2)\ell^2}
 \;.
\end{equation}
A $(r=\tau,t=x)$-projection diagram  can be found in Figure~\ref{F6XI12.1x}.%

No non-trivial, periodic,  time-symmetric ($K_{ij}=0$) spacelike hypersurfaces occur in all space-times above.
Periodic spacelike hypersurfaces with $K_{ij}\not\equiv 0$ arise, but a Hamiltonian analysis of initial data asymptotic to such hypersurfaces goes beyond the scope of this work.

From now on we assume that $f$ has positive zeros.

 \subsection{Spheres and naked singularities}
  \label{ss6I13.2}

\begin{figure}
\begin{center}
{\includegraphics[scale=.85,angle=0]{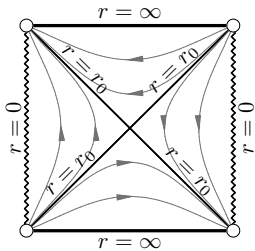}}
\caption{The $(t,r)$-projection diagram for a maximal extension of the Birmingham metrics with $m<0$, $\beta\in \R$, or  $m= 0$ and $\beta=1$, with $r_0$ defined by the condition $f(r_0)=0$. The set $\{r=0\}$ is a singularity unless the metric is the de Sitter metric ($\mathring M= S^{n-1}$ and $m=0$), or a suitable quotient thereof so that $\{r=0\}$ corresponds to a center of (possibly local) rotational symmetry. { }}
\label{F7I13.1}
\end{center}
\end{figure}

Assuming that $m=0$ but $\beta\ne 0$, we must have $\beta=1$ in view of our hypothesis that $f$ has positive zeros. For $r\ge 0$ the function $f$ has exactly one zero,  $r=\ell$.
The boundaries $\{r=0\}$  correspond either to regular
centers of symmetry, in which case the level sets of $t$  are $S^n$'s or their quotients, or
to conical singularities. See Figure~\ref{F7I13.1}.

If $m<0$ the function $f:(0,\infty)\to \R$ is monotonously decreasing, tending to minus infinity as $r$ tends to zero,
where a naked singularity occurs, and to minus infinity when $r$ tends to $\infty$, hence $f$ has then precisely one zero.  The $(t,r)$-projection diagram can be seen again in Figure~\ref{F7I13.1}.

No spatially periodic time-symmetric spacelike hypersurfaces occur in the space-times above.

 \subsection{Spatially periodic time-symmetric initial data}
  \label{ss7I13.11}

\begin{figure}
\begin{center}
{\includegraphics[scale=.85,angle=0]{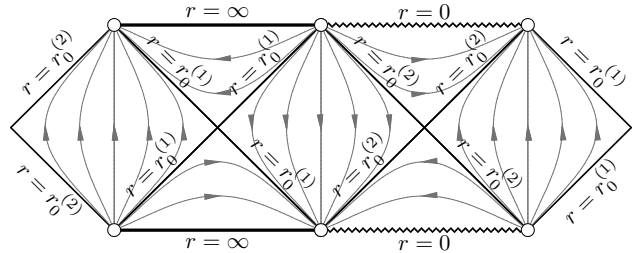}}
\caption{The $(t,r)$-projection diagram for suitably extended Birmingham metrics with exactly two first-order zeros of $f$. The symbols $r_0^{(a)}$, $a=1,2$, denote zeros of $f$.}
\label{F6XI12.1}
\end{center}
\end{figure}

We continue with  the remaining cases, that is, $f$ having zeros and $m> 0$. (When $\beta=1$ this implies $0 <  {m}  \le \frac1n\left(1-\frac2n\right)^{\frac{n}2-1}{\ell^{n-2}}$.)
The function
$f:(0,\infty)\to \R$ is then concave and thus has precisely two first order zeros, except when $m$ attains its maximal allowed value, a case already discussed (see  \eq{6I13.3}).
A projection diagram for a maximal extension of the space-time, for the cases with two-first-order zeros, is provided by  Figure~\ref{F6XI12.1}. The level sets of $t$ within each of the diamonds in that figure can be smoothly continued across the bifurcation surfaces of the Killing horizons to smooth spatially-periodic Cauchy surfaces.

\begin{figure}
\begin{center}
{\includegraphics[scale=0.5,angle=0]{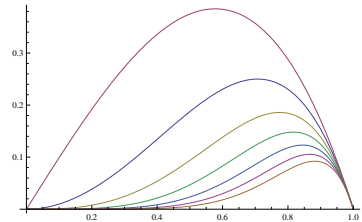}}
\caption{The polynomials $W_n$ for $3\le n \le 9$.}
\label{F13IV13.1}
\end{center}
\end{figure}

Observe that for $\beta=1$ and $0 < \frac{m}{\ell^{n-2}} < \frac1n\left(1-\frac2n\right)^{\frac{n}2-1}$ the roots $r_0^{(a)}$, $a=1,2$, satisfy
\begin{equation}\label{13IV13.1}
 r_0^{(a)} \in ( 0,\ell)
 \;.
\end{equation}
To see this, note that the equation
$\displaystyle f(r_0) = 1 - \frac{2m}{r_0^{n-2}} - \frac{r_0^2}{\ell^2}=0$  is equivalent to
$$
 W_n(x):= x^{n-2}(1-x)(1+x)=\frac{2m}{\ell^{n-2}}
 \; ,
 $$
where $x = {r_0}/\ell$.
The polynomials $W_n  $ are positive precisely on $(0,1)$, which implies the result. Compare Figure~\ref{F13IV13.1}.

\subsection{Killing horizons}
\label{ss28XII12.1}

The locations of Killing horizons of the Birmingham metrics are defined, in space-dimension $n$, by the condition
$$
 f(r_0) = \beta - \frac{2m}{r_0^{n-2}} - \frac{r_0^2}{\ell^2}=0
  \; .
$$
Thus, variations of the metric on the horizons satisfy
\begin{equation}\label{grawit-pow}
0= \delta f|_{r=r_0}= \left. \left[ (\partial_r f) \delta r -\frac2{r^{n-2}}\delta m \right] \right|_{r=r_0}
  \;;
\end{equation}
equivalently
\begin{eqnarray}
  \nonumber
 \delta m
  &= & \frac1{2(n-1)} (\partial_r f) \delta ( r^{n-1})
\\
 & = &
  \frac1{(n-1)\sigma_{n-1}} \frac{(\partial_r f)}2 \Big|_{r=r_0} \delta A  \; ,
 \label{grawit-pow2}
\end{eqnarray}
where $  r^{n-1}\sigma_{n-1}$ is the $\mathring h$-volume  of the cross-section of the horizon.

Let us check  that $\kappa := \frac{(\partial_r f)}2\Big|_{r=r_0}$
coincides with the surface gravity of the horizon, defined through the usual formula
\begin{equation}
 \label{KKK} \nabla_K K = -\kappa K
  \; ,
\end{equation}
where $K$ is the Killing vector field which is null on the horizon.
For this, we rewrite the spacetime metric (\ref{6XI12.4}) in the familiar form
 \[
  g= -f \rd u^2 -2 \rd u \rd r +r^2 \mathring h
  \; ,
   \]
where $\rd u = \rd t - f^{-1} \rd r$.
The Killing field $K= \partial_ { u}=\partial_t$ is indeed tangent to the horizon and null on it.
Formula (\ref{KKK}) implies that
\begin{equation}\label{G-uuu}
    \kappa=-\Gamma^u_{uu}=-\frac12 g^{u\lambda}(2g_{\lambda u,u}-g_{uu,\lambda}) \; .
\end{equation}
The inverse metric equals
\[ g^{\sharp}= -2 \frac\partial{\partial u}\frac\partial{\partial r}+f \left(\frac\partial{\partial r}\right)^2
 +r^{-2} \mathring h^{\sharp}
  \; ,
\]
whence  $g^{u\lambda} = -\delta^\lambda_r$, and
\[
    \kappa=-\frac12 g_{uu,r} = \frac{(\partial_r f)}2\Big|_{r=r_0} \; ,
\]
as claimed. We conclude that on Killing horizons it holds that
\begin{equation}\label{grawit-pow3}
 \delta m =
\frac1{(n-1)\sigma_{n-1}} \kappa \Big|_{r=r_0} \delta A  \;.
\end{equation}

\subsection{Singularities}

Consider a metric of the form
$$
 g=-e^{-2\chi(\tau)} d\tau^2 +  e^{2\chi(\tau)} dx^2 + \tau^2 \mathring h
 \;,
$$
with $\mathring h$ as before. For $A=1,\ldots,n$ let $\ztheta^A$ be an ON-coframe for $\mathring h$,
$$
 \mathring h = \sum_{A=1}^{n-1} \ztheta^A \otimes \ztheta^ A
 \; ,
$$
and let $\mathring \omega _{AB}$ and $\mathring \Omega_{AB}$ be the associated connection and curvature forms, as in the Cartan structure equations:
\beaa
 0 & = & d\ztheta^A + \mathring \omega^A{}_B \wedge \ztheta^B
  \; ,
\\
 \mathring \Omega^A{}_{ B} & = & d \mathring \omega^A{}_{ B}+ \mathring \omega^A{}_{ C}\wedge
  \mathring \omega^C{}_B
  \;.
\end{eqnarray*}
Let $\theta^ \mu$ be the following $g$-ON coframe:
$$
 \theta^0 = e^{-\chi } d\tau
 \; ,
  \quad
 \theta^A = \tau \ztheta ^A
 \; ,
  \quad
 \theta^n = e^\chi dx
 \;.
$$
The condition of vanishing of torsion is solved by setting
\beaa
&
 \omega^n{}_A  = 0
 \; ,
\quad
 \omega^n{}_0  = e^{\chi} \dot \chi \,\theta^n = \frac 12  \dot {( e^{2\chi}) } dx
 \; ,
\\ &
 \omega^A{}_0  = e^{\chi} \ztheta^A
 \; ,
\quad
 \omega^A{}_B  = \mathring
 \omega^A{}_B
 \;.
\end{eqnarray*}
This gives the following curvature two-forms:
%
\beaa
&
 \Omega^0{}_{ n}
=
   \frac 12 \ddot{(e^{2\chi})} \delta^0_{[\mu} g_{\nu] n} \theta^\mu \wedge \theta^\nu
 \; ,
\\
& \Omega^0{}_{ A}
 =
  \frac 12 \dot { (e^{2\chi})} \tau^{-1}\delta^0_{[\mu} g_{\nu] A} \theta^\mu \wedge \theta^\nu
 \; ,
&
\\
&
 \Omega^n{}_{ A}
  =
  \frac 12 \dot { (e^{2\chi})} \tau^{-1}\delta^n_{[\mu} g_{\nu] A} \theta^\mu \wedge \theta^\nu
 \; ,
 \\ &
 \Omega^A{}_{ B}
 =
   \frac 12 \tau^{-2} (\mathring \Omega^A{}_{ B CD} + 2 e^{2\chi} \delta{}^A _{ [C}\delta_{D]B}) \,\theta^C \wedge \theta^D
 \;.
 &
\end{eqnarray*}
Suppose that $g$ is a Birmingham metric with  $m=0$, thus
$$
 e^{2\chi} = -\beta + \frac {\tau^2}{\ell^2}
$$
for a constant $\beta$,
then
$$
  \frac 12 \ddot{(e^{2\chi})} = \frac 12 \dot { (e^{2\chi})} \tau^{-1} = \tau^{-2}(e^{2\chi} + \beta) = \frac 1 {\ell^2}
  \;.
$$
If $\mathring h$ is a space-form, with
$$
 \mathring \Omega^A{}_{BCD } =   2
 \beta \delta^A_{ [C}\delta_{D]B}
 \; ,
$$
consistently with \eq{4I13.4},
we obtain
$$
 R_{\mu\nu\rho\sigma} = \frac 2 {\ell^2} g_{\mu[\rho}g_{\sigma]\nu}
 \;.
$$
If, however,  $\mathring h$ is \emph{not} a space-form, we have
$$
 \mathring \Omega^A{}_{BCD } =   2
 \beta \delta^A_{ [C}\delta_{D]B} + r^A{}_{BCD }
 \; ,
$$
for some non-identically vanishing tensor $r^A{}_{BCD }$, with all traces zero. Hence
$$
 R_{\mu\nu\rho\sigma} = \frac 2 {\ell^2} g_{\mu[\rho}g_{\sigma]\nu} + \tau^{-2} r_{\mu\nu\rho\sigma}
 \; ,
$$
where the functions $r_{\mu\nu\rho\sigma}$ are $\tau$-independent in the current frame, and vanish whenever one
of the indices is $0$ or $n$.
This gives
\beaa
R^ {\mu\nu\rho\sigma} R_{\mu\nu\rho\sigma}
 &= &
   \frac {2 n (n+1)}{\ell^4} +  r^ {\mu\nu\rho\sigma} r_{\mu\nu\rho\sigma}
\\
 &= &
   \frac {2 n (n+1)}{\ell^4} +  \tau^{-4}  \sum_{A,B,C,D=1}^{n-1}
    (r_{ABCD}) ^2
 \; ,
\end{eqnarray*}
which is singular at $\tau=0$.

\section{A control-response calculation}
 \label{s17XII12.11}

To give our considerations  a precise Hamiltonian meaning we need to explicit the family of metrics  considered, as well as the time parameter with respect to which the Hamiltonian will be determined.
The latter is closely related
to a choice of the lapse function.

Here we will consider two distinct settings: a) a boundary $\partial V$ at finite distance with prescribed induced metric there,
 and b) a family of metrics which asymptote, along the asymptotically periodic ends, to Birmingham metrics.

At the boundary, or asymptotically, we make the following choice of
the lapse function
\begin{equation}\label{27II13.1}
     \left.\frac  k\nu \frac {\nu_0}{k_0}\right|_{\partial V} =1 \ \mbox{ or } \  \frac  k\nu \frac {\nu_0}{k_0}  \to 1 \;;
\end{equation}
as already mentioned, this corresponds to a choice of the boundary time, or asymptotic time.
The choice is motivated by the fact that \eq{27II13.1} holds for all metrics in the Birmingham family, see   \eq{27II13.2}-\eq{27II13.3} below.

In the case of a boundary at finite distance, we choose an $(n-1)$-dimensional metric $r^2 \mathring h$   on $\partial V$, as in \eq{6XI12.4}, and consider \emph{the collection of all initial metrics which induce} $r^2 \mathring h$ on $\partial V$.

In the asymptotic case, we choose a compact Riemannian Einstein manifold $(\mathring M, \mathring h)$ and consider \emph{the collection of all metrics which asymptote to the associated Birmingham solutions} along the cylindrical end.

It should be mentioned that the definition of a phase space requires describing also the space of canonical momenta. In the finite-boundary case this issue will be ignored in this work. Concerning asymptotically cylindrical metrics, we will only consider asymptotically vanishing extrinsic curvature tensors $K_{ij}$. We plan to return to asymptotically periodic tensors $K_{ij}$  in future work.

In view of \eq{27II13.1}, when ${\bf Q}_A=0$ and $ \stackrel{\perp}{{Q}}\!{^{AB}}$ is pure trace at $\partial V$,
it is useful (using \eq{1III13.2} and \eq{1III13.3}) to rewrite  the boundary form (\ref{1III13.4}) as
\begin{eqnarray}
 \nonumber
 \lefteqn{
    2 \nu \delta {\bf Q}
- 2 \nu^A \delta {\bf Q}_A
+ \stackrel{\perp}{{Q}}\!{^{AB}}
 \delta g_{AB}
 }\phantom{xx}
 &&
  \\  &&
   =
    \delta(\lambda \nu k) + \lambda \nu^2 \delta (\frac  k\nu ) +
  \nu \big(2  a   - \frac{n-3}{n-1} k \big)\delta \lambda
  \nonumber
\\
   &&
     =
     \delta \left[ \lambda\left(  \nu k - \nu_0 k_0 \right)\right] + \lambda \nu^2 \frac {k_0}{\nu_0} \delta (\frac  k\nu \frac {\nu_0}{k_0})
   + \psiHDS  \; ,
\end{eqnarray}
where
\begin{equation}\label{Psi}
   \psiHDS  :=\delta(\lambda \nu_0 k_0)
   -
   \lambda \nu^2 \frac {k_0}{\nu_0} \frac  k\nu \delta ( \frac {\nu_0}{k_0})+
 \nu\big(2   a   - \frac{n-3}{n-1} k \big) \delta \lambda \; ,
\end{equation}
whereas $k_0 $ and $\nu_0$ are the corresponding quantities calculated on a ``reference configuration'' corresponding to $m=m_0$. For boundaryless configurations with
$$
 {\bf Q}_A\to_{x\to\infty} 0
 \; ,
$$
the above equalities should be understood in the limit $x\to\infty$.

For the Birmingham metrics we have:
\begin{equation}\label{27II13.2}
     f = \beta- \frac{2m}{\phi^{n-2}} - \frac{\phi^2}{\ell^2} \ \ \ ; \ \
     f_0 = \beta - \frac{2m_0}{\phi^{n-2}}-  \frac{\phi^2}{\ell^2} \; ,
\end{equation}
\be
    \nu = \sqrt{f}  \ \ \ ; \ \  \nu_0 = \sqrt{f_0} \; ,
\end{equation}
\be
    k = -\frac {n-1}\phi \sqrt{f} \ \ \ ; \ \
    k_0 = -\frac  {n-1}\phi \sqrt{f_0} \; ,
\end{equation}
\begin{equation}\label{27II13.3}
    \lambda = \phi^{n-1} \sqrt {\det \mathring h} \ \ \ ; \ \
    \nu a = \frac {(n-2)m}{\phi^{n-1}} - \frac \phi{\ell^2} \; .
\end{equation}
This  implies that $\psiHDS $ vanishes identically on $\partial V$ with the above boundary conditions, so that the entire boundary form reduces to
\begin{eqnarray}
 \nonumber
  \lefteqn{
    (  2 n \delta {\bf Q}
- 2 n^A \delta {\bf Q}_A
+ \stackrel{\perp}{{Q}}\!{^{AB}}
 \delta g_{AB} )
 }
  \phantom{xxx}
  &&
\\
  &&
   =
 \delta \left[ \lambda\left(  \nu k - \nu_0 k_0 \right)\right] + \lambda \nu^2 \frac {k_0}{\nu_0} \delta (\frac  k\nu \frac {\nu_0}{k_0}) \; .
 \label{bound-reduce}
\end{eqnarray}
The last term vanishes because of the time gauge \eq{27II13.1},
whereas the first term represents the variation of mass: Indeed, for all Birmingham metrics we have
\begin{equation}\label{energy-fin}
   \lambda\left(  \nu k - \nu_0 k_0 \right) = 2(n-1)(m - m_0)
    \sqrt{\det \mathring h} \; .
\end{equation}
Hence
\begin{equation}\label{16XII12.8a}
  \int_{x=x_0} \!\!\! \lambda(k \nu - k_0\nu_0) +2(n-1) |\partial V|_{\mathring h} m_0 =  2(n-1) |\partial V|_{\mathring h} m
 \; ,
\end{equation}
when the boundary data on $\partial V$ are as above and where, as before, $|\partial V|_{\mathring h}$ denotes the volume of $\partial V$ in the metric $ \mathring h$. In particular the integrand is independent of $x_0$.
Similarly,
\begin{eqnarray}
 \nonumber
 \lefteqn{
 \lim_{x_0 \to
 \infty} \int_{x=x_0} \!\!\! \lambda(k \nu - k_0\nu_0) +2(n-1)|\partial V|_{\mathring h} m_0
  }
  &&
\\
 &&  \phantom{xxxxxxxxxxxxxxxxxx}
 = 2m(n-1)|\partial V|_{\mathring h} \; ,
  \label{16XII12.8b}
\end{eqnarray}
along each asymptotically periodic end.

\section{The Yamabe equation on cylinders}
 \label{ss17XII12.1}

In this section we relate the parameter $m$ appearing in the Schwarzschild-de Sitter metrics to a Hamiltonian for
the spherically-symmetric Yamabe equation. The reader should note that the   Hamiltonian  here is a Hamiltonian
for the dynamics in $x$, not to be confused with that for the dynamics in time, as used elsewhere in this work.

Let
\newcommand{\phiHDS}{\varphi}
\begin{equation}\label{6XI12.0}
  g_{ij}= \phiHDS^{\frac 4 {n-2}} \tilde g_{ij}
  \;.
\end{equation}
Recall the vacuum Lichnerowicz equation with cosmological constant $\Lambda$, in space-dimension $n$,
\begin{equation}\label{conf214}
 \Delta_\tg \phiHDS - \frac {n-2}{4(n-1)}\tR \phiHDS =-
 \mytsigmasquare  \phiHDS^{(2-3n)/(n-2)} + \tbeta\phiHDS^{\frac{n+2}{n-2}}
 \; ,
\end{equation}
where
\begin{equation}\label{conf216}
 \mytsigmasquare :=\frac
 {n-2}{4(n-1)}|\tL|_\tg^2\; ,\quad  \tbeta:= \frac{n-2}{4n}\tau^2 - \frac
 {n-2}{2(n-1)}\Lambda
 \;.
\end{equation}
Here $\tL_{ij}$ is  $\tg$-transverse traceless, and $\tau$ is the trace of the extrinsic curvature tensor $\tau = g^{ij}K_{ij}$, assumed to be constant, with $K_{ij}$  obtained from $\tL_{ij}$ by the usual formula.

Suppose that
\begin{equation}\label{6XI12.1}
 \tg= dx^2 + \mathring h
 \; ,
\end{equation}
where $\mathring h$ is as in \eq{6XI12.4}. We then have $\tR = \mathring R$, and when $\tau$ is a constant we can seek an $x^A$-independent solution of \eq{conf214} with $\tL_{ij}{}=0$:
\begin{equation}\label{6XI12.2}
   \frac{d^2\phiHDS}{dx^2}  - \frac {n-2}{4(n-1)}\zR \phiHDS =  \tbeta\phiHDS^{\frac{n+2}{n-2}}
 \;.
\end{equation}

Equation \eq{6XI12.2} has a usual first integral: setting
\begin{equation}\label{6XI12.3}
   H = \frac 12 \left(\frac{d\phiHDS}{dx}\right)^{2}  - \frac {n-2}{8(n-1)}\zR \phiHDS^2  -   \frac{(n-2)\tbeta}{2n}\phiHDS^{\frac{2n}{n-2}}
 \; ,
\end{equation}
we have
$$
 \frac{d H}{dx}=0
 \;.
$$

We apply the above to the Birmingham metrics with $f\ge 0$; as discussed in Appendix~\ref{s4I13.1}, the metrics with $f\le 0$ do not occur  as asymptotic models   in our context.
We only consider regions,  where $f>0$, the final formulae remain valid at $f=0$ by continuity.

The field of unit normals $N$ to the static slices $t=\const$, which we denote by $\hyp_t$, is given by
\begin{equation}\label{6XI12.5b}
 N = \frac 1 {\sqrt{f}} \partial_t
 \;.
\end{equation}
For those slices we have $\tau=0$.

The volume form $d\mu_{\mathring M}$ on the submanifolds of constant $t$ and $x$ reads
\begin{equation}\label{6XI12.6}
 d\mu_{\mathring M} = \lambda d\mathring \mu_{\mathring M}
 \; ,
 \ \mbox{with} \
 d\mathring \mu_{{\mathring M}} = \sqrt{\det \mathring h_{AB}} d^{n-1} x
 \; ,
\end{equation}
and where
\begin{equation}\label{6XI12.7}
 \lambda = r^{n-1} = \phiHDS^{\frac {2(n-1)}{n-2}}
 \; ,
\end{equation}
with $\phiHDS$ as in \eq{6XI12.0}-\eq{conf214}:
\begin{equation}\label{6XO12.8}
\gamma:= \frac{dr^2}{f} + r^2 \mathring h = \phiHDS^{\frac{4}{n-2}}(dx^2 + \mathring h)
 \;.
\end{equation}
The last equation implies
\beal{6XI12.9}
 & \displaystyle
  \phiHDS = r^{\frac {n-2}2}
 \; ,
 \quad
 \frac{dr}{dx}= \phiHDS^{\frac 2 {n-2}}\sqrt{f}
 = r\sqrt{f}
 \; ,
 &
\\
 & \displaystyle
 \sqrt{\det \gamma}= \frac{r^{n-1}\sqrt{\det \mathring h}}{\sqrt{f}}
 \ \mbox{or} \  \phiHDS^{\frac {2n}{n-2}}\sqrt{\det \mathring h}
 \;.
 &\label{6XI12.9x}
\end{eqnarray}

Let $\mathbf m$ denote the field of unit normals to the level sets of $r$ within $\hyp_t$, and let $k$ denote the extrinsic curvature, within $\hyp_t$, of those level sets. We have

\beal{6XI12.10}
 {\mathbf m}  &= & \sqrt{f} \partial_r = \phiHDS^{- \frac 2{n-2}}\partial_x
 \; ,
\\
 k &= &  \frac 1 {\sqrt{\det \gamma}} \partial_r( {\sqrt{\det \gamma}} \,{\mathbf m}^r)
 \frac{(n-1)}r \sqrt f
 \nonumber
\\
 &  =&
  \phiHDS^{ - \frac {2n} {n-2}}\partial_x (  \phiHDS^{ \frac{2(n-1)}{n-2}} )  =  \frac{2(n-1)}{n-2} \phiHDS^{ - \frac {n}{n-2} }\partial_x \phiHDS
  \;.
   \phantom{xxx}\label{6XI12.11}
\end{eqnarray}
It follows that
\begin{equation}\label{6XI12.12}
 \partial_x \phiHDS =
 \frac{(n-2)\phiHDS^{  \frac {n}{n-2}}} {2r}\sqrt f =
  \frac{(n-2)}{2} {r^{  \frac   {n-2} 2}} \sqrt f
 \; ,
\end{equation}
and that the constant of motion $H$ of \eq{6XI12.3} equals
\bean
   H
    & = &
      \frac 12 \left(\frac{d\phiHDS}{dx}\right)^{2}  - \frac {n-2}{8(n-1)}\zR \phiHDS^2  -   \frac{(n-2)\tbeta}{2n}\phiHDS^{\frac{2n}{n-2}}
\\
  &= &  \frac {(n-2)^{2}}8
  r^{    {n-2}  }  \left[ f   - \frac{\zR}{(n-2)(n-1)}   -   \frac{4\tbeta}{n(n-2)}r^{2} \right]
  \nonumber
\\
 & = &  - \frac {(n-2)^{2}}4 m
 \; ,\label{6XI12.13}
\end{eqnarray}
provided that
$$
  \frac {(n-2)^{2}}{8\ell^2} = - \frac{(n-2)\tbeta}{2n}
   =   \frac{(n-2)}{2n}\times \frac
 {(n-2)}{2(n-1)}\Lambda
  \; ,
$$
which will be the case if
\begin{equation}\label{17XII12.21}
 \frac 1 {\ell^2} = \frac {2\Lambda}{n(n-1)}
 \;.
\end{equation}

\section{Einstein equations in $n+1$ dimensions}
 \label{s3I13.1}

Hamiltonian dynamics is usually derived from a Lagrangean. The latter is determined by the equations of the theory
up to a multiplicative constant. One therefore needs a prescription which determines this constant. For this, we decree
that for a point-particle of rest mass $m_0$ moving on a timelike curve $\Gamma$ the Lagrangean is, independently of dimension,
\begin{equation}\label{3I13.6}
 \mcL_{m_0} = -m_0 \int_\R \sqrt{g(\dot \Gamma,\dot \Gamma)}dt
 \;.
\end{equation}
Equivalently, the energy-momentum tensor $T_{\mu\nu}:=\partial \mcL_{m_0} /\partial g^{\mu\nu}$ of such a particle is
$$
 T_{\mu\nu} = m_0 u_\mu u_\nu \,\delta_\Gamma
 \; ,
$$
where $\delta_\Gamma $ is the distribution acting on functions as
$$
 \langle \delta_\Gamma,f\rangle = \int_\R (f\circ \Gamma) (t)\sqrt{|g(\dot \Gamma,\dot \Gamma)|}dt
 \;.
$$

The  Einstein equations in $n+1$ dimensions, which we write in the form
\begin{equation}\label{3I13.7}
 G_{\mu\nu} = \gamma T_{\mu\nu}
 \; ,
\end{equation}
where $\gamma$ is a dimension-dependent constant, are compatible with \eq{3I13.7} if
\begin{equation}\label{3I13.8}
 \mcL = \frac 1 {2\gamma} \int R \,\mu_g -\mcL_{m_0}
  \; .
\end{equation}

We emphasize that the considerations here are \emph{not} supposed to be rigorous. The aim is to give a heuristic
justification of the choice of the constants involved, and the questions of convergence of the integrals, or consistency of the scheme, are completely irrelevant for our purposes.

In order to relate the value of $\gamma$ to physics in $n+1$ dimensions we consider the ``Newtonian limit" of
\eq{3I13.7}:
We assume that the metric is time-independent, and takes the form
$$
 g_{\mu\nu} = \eta_{\mu\nu} + h_{\mu\nu}
 \; ,
$$
where $\eta_{\mu\nu}$ is the Minkowski metric.
We suppose that all expressions quadratic in the $h_{\mu\nu}$'s  and their derivatives can be neglected in the calculations that follow. Taking $T_{\mu\nu}$ of
the form $\rho \delta^0_\mu \delta^0 _\nu$, and a harmonic gauge
$$
 \partial_\mu(\underbrace{ h^{\mu\nu}- \frac 12 \eta^{\alpha\beta} h_{\alpha\beta} \eta^{\mu\nu}}_{=:\hbar ^{\mu\nu}} ) = 0
$$
(with all indices raised and lowered with the Minkowski metric), a standard calculation in which time-derivatives are also neglected leads to
\begin{equation}\label{3I12.1}
 - \frac 12 \Delta_e \hbar_{\mu\nu} =  \rho \delta^0_\mu \delta^0 _\nu
 \; ,
\end{equation}
where $\Delta_e$ is the Laplace operator of the Euclidean metric.

Recall the identity, in space-dimension $n\ge 3$,
$$
 \Delta_e  \frac 1 {r^{n-2} }= -(n-2) \omega_{n-1} \delta_0
 \; ,
$$
where $\omega_d$ denotes the volume of a unit, round $d$-dimensional sphere.
The solution of \eq{3I12.1} for a point distribution with total mass $M$ therefore takes the form
\begin{equation}\label{3I12.2}
 \hbar _{\mu\nu} =  \frac {2 \,\gamma \, M} {  (n-2) \omega_{n-1}r^{n-2} } \delta^0_\mu \delta^0 _\nu
 \;.
\end{equation}

Consider an approximate geodesic of the form $(t,\vec x (t))$. Assuming that all terms quadratic in $ \dot{\vec x}$ and
its derivatives can be neglected, the coordinate acceleration vector $\vec a$ equals
$$
 a^k = \ddot x^k \approx -\Gamma^k_{00} \approx \frac 12 \partial_k h_{00} =: - \partial_k \varphi
 \; ,
$$
where $\varphi$ is the ``Newtonian" potential.    From \eq{3I12.2} we have
$$
 h_{00} =  \frac {2 \,\gamma \, M} {  (n-1) \omega_{n-1}r^{n-2} }
 \; ,
$$
leading to
$$
 \varphi =  -\frac { \gamma \, M} {  (n-1) \omega_{n-1}r^{n-2} }
 \;.
$$
This makes it clear how $\gamma$ is related to the $(n+1)$-dimensional Newton constant $G_n$:
\begin{align} \nonumber
\vec F \equiv  m_0\vec a = - G_n  m_0  M \frac {\vec x } {  r^{n-1} } = - m_0 \nabla \varphi
 \\
  \quad
  \Longleftrightarrow
  \quad
  \gamma = \frac {(n-1)\omega_{n-1}}{n-2} G_n
  \; . \label{3I13.11}
\end{align}

{\noindent \sc Acknowledgements} We are grateful to Christa Raphaela \"Olz for providing the figures,
and to Bobby Beig for useful {discussions} and bibliographical advice. Supported in part by Narodowe Centrum Nauki (Poland) under the grant DEC-2011/03/B/ST1/02625
and the Austrian Science Fund (FWF) under project   P 23719-N16.  JJ and JK wish to thank the Erwin Schr\"odinger Institute, Vienna, for hospitality and support during part of work on this paper.

\bibliography{ChruscielJezierskiKijowski-minimal}

\def\polhk#1{\setbox0=\hbox{#1}{\ooalign{\hidewidth
  \lower1.5ex\hbox{`}\hidewidth\crcr\unhbox0}}} \def\cprime{$'$}
  \def\cprime{$'$}
\begin{thebibliography}{23}%
\makeatletter
\providecommand \@ifxundefined [1]{%
 \@ifx{#1\undefined}
}%
\providecommand \@ifnum [1]{%
 \ifnum #1\expandafter \@firstoftwo
 \else \expandafter \@secondoftwo
 \fi
}%
\providecommand \@ifx [1]{%
 \ifx #1\expandafter \@firstoftwo
 \else \expandafter \@secondoftwo
 \fi
}%
\providecommand \natexlab [1]{#1}%
\providecommand \enquote  [1]{``#1''}%
\providecommand \bibnamefont  [1]{#1}%
\providecommand \bibfnamefont [1]{#1}%
\providecommand \citenamefont [1]{#1}%
\providecommand \href@noop [0]{\@secondoftwo}%
\providecommand \href [0]{\begingroup \@sanitize@url \@href}%
\providecommand \@href[1]{\@@startlink{#1}\@@href}%
\providecommand \@@href[1]{\endgroup#1\@@endlink}%
\providecommand \@sanitize@url [0]{\catcode `\\12\catcode `\$12\catcode
  `\&12\catcode `\#12\catcode `\^12\catcode `\_12\catcode `\%12\relax}%
\providecommand \@@startlink[1]{}%
\providecommand \@@endlink[0]{}%
\providecommand \url  [0]{\begingroup\@sanitize@url \@url }%
\providecommand \@url [1]{\endgroup\@href {#1}{\urlprefix }}%
\providecommand \urlprefix  [0]{URL }%
\providecommand \Eprint [0]{\href }%
\providecommand \doibase [0]{http://dx.doi.org/}%
\providecommand \selectlanguage [0]{\@gobble}%
\providecommand \bibinfo  [0]{\@secondoftwo}%
\providecommand \bibfield  [0]{\@secondoftwo}%
\providecommand \translation [1]{[#1]}%
\providecommand \BibitemOpen [0]{}%
\providecommand \bibitemStop [0]{}%
\providecommand \bibitemNoStop [0]{.\EOS\space}%
\providecommand \EOS [0]{\spacefactor3000\relax}%
\providecommand \BibitemShut  [1]{\csname bibitem#1\endcsname}%
\let\auto@bib@innerbib\@empty
\bibitem [{\citenamefont {Chru\'{s}ciel}\ and\ \citenamefont
  {Pollack}(2008)}]{ChPollack}%
  \BibitemOpen
  \bibfield  {author} {\bibinfo {author} {\bibfnamefont {P.}~\bibnamefont
  {Chru\'{s}ciel}}\ and\ \bibinfo {author} {\bibfnamefont {D.}~\bibnamefont
  {Pollack}},\ }\href {\doibase 10.1007/s00023-008-0368-6} {\bibfield
  {journal} {\bibinfo  {journal} {Ann.\ Henri Poincar\'e}\ }\textbf {\bibinfo
  {volume} {9}},\ \bibinfo {pages} {639} (\bibinfo {year} {2008})},\ \bibinfo
  {note} {arXiv:0710.3365 [gr-qc]}\BibitemShut {NoStop}%
\bibitem [{\citenamefont {Chru\'{s}ciel}\ \emph {et~al.}(2008)\citenamefont
  {Chru\'{s}ciel}, \citenamefont {Pacard},\ and\ \citenamefont
  {Pollack}}]{CPP}%
  \BibitemOpen
  \bibfield  {author} {\bibinfo {author} {\bibfnamefont {P.}~\bibnamefont
  {Chru\'{s}ciel}}, \bibinfo {author} {\bibfnamefont {F.}~\bibnamefont
  {Pacard}}, \ and\ \bibinfo {author} {\bibfnamefont {D.}~\bibnamefont
  {Pollack}},\ }\href@noop {} {\bibfield  {journal} {\bibinfo  {journal}
  {Math.\ Res.\ Lett.}\ }\textbf {\bibinfo {volume} {16}},\ \bibinfo {pages}
  {157} (\bibinfo {year} {2008})},\ \bibinfo {note} {arXiv:0803.1817
  [gr-qc]}\BibitemShut {NoStop}%
\bibitem [{\citenamefont {Cortier}(2012)}]{CortierKdS}%
  \BibitemOpen
  \bibfield  {author} {\bibinfo {author} {\bibfnamefont {J.}~\bibnamefont
  {Cortier}},\ }\href@noop {} {\  (\bibinfo {year} {2012})},\ \Eprint
  {http://arxiv.org/abs/1202.3688} {arXiv:1202.3688 [gr-qc]} \BibitemShut
  {NoStop}%
\bibitem [{\citenamefont {Chru\'{s}ciel}\ \emph
  {et~al.}(2012{\natexlab{a}})\citenamefont {Chru\'{s}ciel}, \citenamefont
  {Mazzeo},\ and\ \citenamefont {Pocchiola}}]{CMP}%
  \BibitemOpen
  \bibfield  {author} {\bibinfo {author} {\bibfnamefont {P.}~\bibnamefont
  {Chru\'{s}ciel}}, \bibinfo {author} {\bibfnamefont {R.}~\bibnamefont
  {Mazzeo}}, \ and\ \bibinfo {author} {\bibfnamefont {S.}~\bibnamefont
  {Pocchiola}},\ }\href@noop {} {\  (\bibinfo {year} {2012}{\natexlab{a}})},\
  \bibinfo {note} {arXiv:1203.5138 [gr-qc]}\BibitemShut {NoStop}%
\bibitem [{\citenamefont {Chru\'{s}ciel}\ and\ \citenamefont
  {Mazzeo}(2012)}]{CM}%
  \BibitemOpen
  \bibfield  {author} {\bibinfo {author} {\bibfnamefont {P.}~\bibnamefont
  {Chru\'{s}ciel}}\ and\ \bibinfo {author} {\bibfnamefont {R.}~\bibnamefont
  {Mazzeo}},\ }\href@noop {} {\  (\bibinfo {year} {2012})},\ \Eprint
  {http://arxiv.org/abs/1201.4937} {arXiv:1201.4937 [gr-qc]} \BibitemShut
  {NoStop}%
\bibitem [{\citenamefont {Waxenegger}\ \emph {et~al.}(2011)\citenamefont
  {Waxenegger}, \citenamefont {Beig},\ and\ \citenamefont {Murchadha}}]{BMW}%
  \BibitemOpen
  \bibfield  {author} {\bibinfo {author} {\bibfnamefont {G.}~\bibnamefont
  {Waxenegger}}, \bibinfo {author} {\bibfnamefont {R.}~\bibnamefont {Beig}}, \
  and\ \bibinfo {author} {\bibfnamefont {N.}~\bibnamefont {Murchadha}},\ }\href
  {\doibase 10.1088/0264-9381/28/24/245002} {\bibfield  {journal} {\bibinfo
  {journal} {Class.\ Quantum Grav.}\ }\textbf {\bibinfo {volume} {28}},\
  \bibinfo {pages} {245002, pp.~15} (\bibinfo {year} {2011})},\ \Eprint
  {http://arxiv.org/abs/1107.3083} {arXiv:1107.3083 [gr-qc]} \BibitemShut
  {NoStop}%
\bibitem [{\citenamefont {Cl{\'e}ment}(2010)}]{GabachClement}%
  \BibitemOpen
  \bibfield  {author} {\bibinfo {author} {\bibfnamefont {M.~G.}\ \bibnamefont
  {Cl{\'e}ment}},\ }\href {\doibase 10.1088/0264-9381/27/12/125010} {\bibfield
  {journal} {\bibinfo  {journal} {Class.\ Quantum Grav.}\ }\textbf {\bibinfo
  {volume} {27}},\ \bibinfo {pages} {125010} (\bibinfo {year} {2010})},\
  \Eprint {http://arxiv.org/abs/0911.0258} {arXiv:0911.0258 [gr-qc]}
  \BibitemShut {NoStop}%
\bibitem [{\citenamefont {Abbott}\ and\ \citenamefont
  {Deser}(1982)}]{AbbottDeser}%
  \BibitemOpen
  \bibfield  {author} {\bibinfo {author} {\bibfnamefont {L.}~\bibnamefont
  {Abbott}}\ and\ \bibinfo {author} {\bibfnamefont {S.}~\bibnamefont {Deser}},\
  }\href@noop {} {\bibfield  {journal} {\bibinfo  {journal} {Nucl.\ Phys.}\
  }\textbf {\bibinfo {volume} {B195}},\ \bibinfo {pages} {76} (\bibinfo {year}
  {1982})}\BibitemShut {NoStop}%
\bibitem [{\citenamefont {Anninos}(2012)}]{AnninosMusings}%
  \BibitemOpen
  \bibfield  {author} {\bibinfo {author} {\bibfnamefont {D.}~\bibnamefont
  {Anninos}},\ }\href {\doibase 10.1142/S0217751X1230013X} {\bibfield
  {journal} {\bibinfo  {journal} {Internat.\ Jour.\ Modern Phys.\ A}\ }\textbf
  {\bibinfo {volume} {27}},\ \bibinfo {pages} {1230013, pp.~38} (\bibinfo
  {year} {2012})}\BibitemShut {NoStop}%
\bibitem [{\citenamefont {Balasubramanian}\ \emph {et~al.}(2002)\citenamefont
  {Balasubramanian}, \citenamefont {de~Boer},\ and\ \citenamefont
  {Minic}}]{BalaBD}%
  \BibitemOpen
  \bibfield  {author} {\bibinfo {author} {\bibfnamefont {V.}~\bibnamefont
  {Balasubramanian}}, \bibinfo {author} {\bibfnamefont {J.}~\bibnamefont
  {de~Boer}}, \ and\ \bibinfo {author} {\bibfnamefont {D.}~\bibnamefont
  {Minic}},\ }\href {\doibase 10.1103/PhysRevD.65.123508} {\bibfield  {journal}
  {\bibinfo  {journal} {Phys.\ Rev.}\ }\textbf {\bibinfo {volume} {D65}},\
  \bibinfo {pages} {123508, pp.~15} (\bibinfo {year} {2002})},\ \Eprint
  {http://arxiv.org/abs/hep-th/0110108} {arXiv:hep-th/0110108 [hep-th]}
  \BibitemShut {NoStop}%
\bibitem [{\citenamefont {Kijowski}(1997)}]{KijowskiGRG}%
  \BibitemOpen
  \bibfield  {author} {\bibinfo {author} {\bibfnamefont {J.}~\bibnamefont
  {Kijowski}},\ }\href {\doibase 10.1023/A:1010268818255} {\bibfield  {journal}
  {\bibinfo  {journal} {Gen.\ Rel.\ Grav.}\ }\textbf {\bibinfo {volume} {29}},\
  \bibinfo {pages} {307} (\bibinfo {year} {1997})}\BibitemShut {NoStop}%
\bibitem [{\citenamefont {Kijowski}\ and\ \citenamefont
  {Tulczyjew}(1979)}]{KijowskiTulczyjew}%
  \BibitemOpen
  \bibfield  {author} {\bibinfo {author} {\bibfnamefont {J.}~\bibnamefont
  {Kijowski}}\ and\ \bibinfo {author} {\bibfnamefont {W.}~\bibnamefont
  {Tulczyjew}},\ }\href@noop {} {\emph {\bibinfo {title} {A Symplectic
  Framework for Field Theories}}},\ \bibinfo {series} {Lecture Notes in
  Physics}, Vol.\ \bibinfo {volume} {107}\ (\bibinfo  {publisher} {Springer},\
  \bibinfo {address} {New York, Heidelberg, Berlin},\ \bibinfo {year} {1979})\
  pp.\ \bibinfo {pages} {iv+257}\BibitemShut {NoStop}%
\bibitem [{\citenamefont {Chru\'{s}ciel}\ \emph {et~al.}(2001)\citenamefont
  {Chru\'{s}ciel}, \citenamefont {Jezierski},\ and\ \citenamefont
  {Kijowski}}]{CJK}%
  \BibitemOpen
  \bibfield  {author} {\bibinfo {author} {\bibfnamefont {P.}~\bibnamefont
  {Chru\'{s}ciel}}, \bibinfo {author} {\bibfnamefont {J.}~\bibnamefont
  {Jezierski}}, \ and\ \bibinfo {author} {\bibfnamefont {J.}~\bibnamefont
  {Kijowski}},\ }\href@noop {} {\emph {\bibinfo {title} {{H}amiltonian field
  theory in the radiating regime}}},\ \bibinfo {series} {Lect. Notes in
  Physics}, Vol.\ \bibinfo {volume} {m70}\ (\bibinfo  {publisher} {Springer},\
  \bibinfo {address} {Berlin, Heidelberg, New York},\ \bibinfo {year} {2001})\
  pp.\ \bibinfo {pages} {vi+172}\BibitemShut {NoStop}%
\bibitem [{Note1()}]{Note1}%
  \BibitemOpen
  \bibinfo {note} {In this work we consider only asymptotically time-symmetric
  models, and therefore only the solutions discussed in Section~\ref
  {ss7I13.11}: The exactly cylindrical solutions of Appendix~\ref {ss6I13.1}
  are not time-symmetric, while the solutions of Appendix~\ref {ss6I13.2} have
  no asymptotically cylindrical regions.}\BibitemShut {Stop}%
\bibitem [{Note2()}]{Note2}%
  \BibitemOpen
  \bibinfo {note} {Indeed, both extremal values $r_0^{(1)}$ and $r_0^{(2)}$ of
  Figure \ref {F6XI12.1} belong to the interval $[0,\ell ]$, see (\ref
  {13IV13.1}).}\BibitemShut {Stop}%
\bibitem [{\citenamefont {Chru\'{s}ciel}\ \emph {et~al.}(2004)\citenamefont
  {Chru\'{s}ciel}, \citenamefont {Jezierski},\ and\ \citenamefont
  {\L\c{e}ski}}]{CJL}%
  \BibitemOpen
  \bibfield  {author} {\bibinfo {author} {\bibfnamefont {P.}~\bibnamefont
  {Chru\'{s}ciel}}, \bibinfo {author} {\bibfnamefont {J.}~\bibnamefont
  {Jezierski}}, \ and\ \bibinfo {author} {\bibfnamefont {S.}~\bibnamefont
  {\L\c{e}ski}},\ }\href@noop {} {\bibfield  {journal} {\bibinfo  {journal}
  {Adv.\ Theor.\ Math.\ Phys.}\ }\textbf {\bibinfo {volume} {8}},\ \bibinfo
  {pages} {83} (\bibinfo {year} {2004})},\ \bibinfo {note}
  {arXiv:gr-qc/0307109}\BibitemShut {NoStop}%
\bibitem [{\citenamefont {Chru\'{s}ciel}(1985)}]{ChAIHP}%
  \BibitemOpen
  \bibfield  {author} {\bibinfo {author} {\bibfnamefont {P.}~\bibnamefont
  {Chru\'{s}ciel}},\ }\href
  {http://www.numdam.org/item?id=AIHPB_1985__42_3_267_0} {\bibfield  {journal}
  {\bibinfo  {journal} {Ann.\ Inst.\ Henri Poincar\'e}\ }\textbf {\bibinfo
  {volume} {42}},\ \bibinfo {pages} {267} (\bibinfo {year} {1985})}\BibitemShut
  {NoStop}%
\bibitem [{\citenamefont {Beig}\ and\ \citenamefont
  {Murchadha}(1987)}]{BOM:poincare}%
  \BibitemOpen
  \bibfield  {author} {\bibinfo {author} {\bibfnamefont {R.}~\bibnamefont
  {Beig}}\ and\ \bibinfo {author} {\bibfnamefont {N.~{\'O}.}\ \bibnamefont
  {Murchadha}},\ }\href@noop {} {\bibfield  {journal} {\bibinfo  {journal}
  {Ann. Phys.}\ }\textbf {\bibinfo {volume} {174}},\ \bibinfo {pages} {463}
  (\bibinfo {year} {1987})}\BibitemShut {NoStop}%
\bibitem [{\citenamefont {Czuchry}\ \emph {et~al.}(2004)\citenamefont
  {Czuchry}, \citenamefont {Jezierski},\ and\ \citenamefont
  {Kijowski}}]{CJKbhthermo}%
  \BibitemOpen
  \bibfield  {author} {\bibinfo {author} {\bibfnamefont {E.}~\bibnamefont
  {Czuchry}}, \bibinfo {author} {\bibfnamefont {J.}~\bibnamefont {Jezierski}},
  \ and\ \bibinfo {author} {\bibfnamefont {J.}~\bibnamefont {Kijowski}},\
  }\href {\doibase 10.1103/PhysRevD.70.124010} {\bibfield  {journal} {\bibinfo
  {journal} {Phys. Rev. D (3)}\ }\textbf {\bibinfo {volume} {70}},\ \bibinfo
  {pages} {124010, 14} (\bibinfo {year} {2004})},\ \bibinfo {note}
  {arXiv:gr-qc/0412042}\BibitemShut {NoStop}%
\bibitem [{Note3()}]{Note3}%
  \BibitemOpen
  \bibinfo {note} {The formula has been proved in dimension three in~\cite
  {CJKbhthermo}, but remains true in higher dimensions.}\BibitemShut {Stop}%
\bibitem [{\citenamefont {Birmingham}(1999)}]{Birmingham}%
  \BibitemOpen
  \bibfield  {author} {\bibinfo {author} {\bibfnamefont {D.}~\bibnamefont
  {Birmingham}},\ }\href@noop {} {\bibfield  {journal} {\bibinfo  {journal}
  {Class.\ Quantum Grav.}\ }\textbf {\bibinfo {volume} {16}},\ \bibinfo {pages}
  {1197} (\bibinfo {year} {1999})},\ \bibinfo {note}
  {arXiv:hep-th/9808032}\BibitemShut {NoStop}%
\bibitem [{\citenamefont {Chru\'{s}ciel}\ \emph
  {et~al.}(2012{\natexlab{b}})\citenamefont {Chru\'{s}ciel}, \citenamefont
  {\"Olz},\ and\ \citenamefont {Szybka}}]{COS}%
  \BibitemOpen
  \bibfield  {author} {\bibinfo {author} {\bibfnamefont {P.}~\bibnamefont
  {Chru\'{s}ciel}}, \bibinfo {author} {\bibfnamefont {C.}~\bibnamefont
  {\"Olz}}, \ and\ \bibinfo {author} {\bibfnamefont {S.}~\bibnamefont
  {Szybka}},\ }\href@noop {} {\bibfield  {journal} {\bibinfo  {journal} {Phys.\
  Rev.\ D}\ }\textbf {\bibinfo {volume} {86}},\ \bibinfo {pages} {124041,
  pp.~20} (\bibinfo {year} {2012}{\natexlab{b}})},\ \Eprint
  {http://arxiv.org/abs/1211.1718} {arXiv:1211.1718 [gr-qc]} \BibitemShut
  {NoStop}%
\bibitem [{\citenamefont {Hawking}\ and\ \citenamefont {Ellis}(1973)}]{HE}%
  \BibitemOpen
  \bibfield  {author} {\bibinfo {author} {\bibfnamefont {S.}~\bibnamefont
  {Hawking}}\ and\ \bibinfo {author} {\bibfnamefont {G.}~\bibnamefont
  {Ellis}},\ }\href@noop {} {\emph {\bibinfo {title} {The large scale structure
  of space-time}}}\ (\bibinfo  {publisher} {Cambridge University Press},\
  \bibinfo {address} {Cambridge},\ \bibinfo {year} {1973})\ pp.\ \bibinfo
  {pages} {xi+391}\BibitemShut {NoStop}%
\end{thebibliography}%

\end{document}